# Low-temperature thermal expansion of the α and ω phases of titanium

Norimasa Nishiyama[1,*], Atsushi Togo[2], Takuya Sasaki[3], Yoshitaka Matsushita[4]

1: Research Center for Materials Nanoarchitectonics, National Institute for Materials Science, 1-1 Namiki, Tsukuba 305-0044, Japan

2: Center for Basic Research on Materials, National Institute for Materials Science, 1-2-1 Sengen, Tsukuba 305-0047, Japan

3: Department of Materials Physics, Graduate School of Engineering, Nagoya University, Furo-cho, Chikusa-ku, Nagoya, Aichi 464-8603, Japan

4: Advanced Engineering and Services Division, National Institute for Materials Science, 1-1 Namiki, Tsukuba 305-0044, Japan

*Corresponding author: Norimasa Nishiyama

Address: Research Center for Materials Nanoarchitectonics, National Institute for Materials Science, 1-1 Namiki, Tsukuba 305-0044, Japan

Email: nishiyama.norimasa@nims.go.jp

**Abstract**

We investigated the low-temperature thermal expansion of the α phase and the ω phase, a high-pressure polymorph of titanium, by X-ray diffraction measurements from approximately 5 to 300 K, together with first-principles calculations. Single-phase bulk polycrystalline ω-Ti was synthesized at 7.7 GPa and 500 °C and recovered to ambient pressure. The two phases exhibit very similar volumetric thermal expansion. Debye-Grüneisen analysis of the temperature-dependent unit-cell volumes yielded Debye temperatures of 427 ± 5 K and 409 ± 12 K and effective Grüneisen parameters of 1.33 ± 0.04 and 1.23 ± 0.04 for α-Ti and ω-Ti, respectively. In contrast, their axial thermal expansion differs markedly: α-Ti exhibits anisotropic expansion, including negative thermal expansion along the *c* axis below approximately 50 K, whereas ω-Ti exhibits nearly isotropic thermal expansion. First-principles calculations reproduce the volumetric thermal expansion of both phases and capture the overall trends of their axial responses. The calculated elastic constants indicate substantial elastic anisotropy in ω-Ti despite its nearly isotropic thermal expansion. The anisotropic thermal expansion of α-Ti and the nearly isotropic thermal expansion of ω-Ti can be understood in terms of the balance between elastic coupling and temperature-dependent thermal driving forces.

## 1. Introduction

Thermal expansion is a fundamental physical property of solids that reflects the temperature-dependent response of a crystal lattice. In noncubic crystals, this response is generally anisotropic, and the lattice parameters along different crystallographic directions may exhibit distinct temperature dependences. At ambient pressure, titanium adopts the hexagonal close-packed α phase. Although the thermal expansion of polycrystalline α-Ti has been reported over wide temperature ranges in several studies [1-4], experimental studies of linear thermal expansion along the *a* and *c* axes remain limited. Nizhankovskii et al. [5] measured the thermal expansion of single-crystal α-Ti along both the a and c axes at low temperatures. The authors reported negative thermal expansion along the *c* axis at temperatures below 165 K. Some first-principles studies have reproduced the occurrence of negative *c*-axis thermal expansion and proposed mechanisms for this anisotropic behavior [6,7]. However, the calculated thermal-expansion coefficients do not quantitatively agree with the reported experimental values. A later conference abstract by Hope and Kodess [8] reported XRD measurements of the lattice parameters of α-Ti at several temperatures between 10 and 290 K and noted that their results were inconsistent with the previously reported low-temperature negative thermal expansion along the *c* axis. Nevertheless, detailed experimental data for both the *a* and *c* axes in the low-temperature range where negative *c*-axis expansion occurs remain very limited. An independent diffraction-based determination of the temperature-dependent *a* and *c* lattice parameters is therefore valuable for reinvestigating the low-temperature thermal expansion behavior of α-Ti and for comparison with theoretical calculations.

Titanium forms the hexagonal ω phase under high pressure [9,10]. At ambient pressure, metastable ω precipitates are commonly observed in β-titanium alloys [11,12]. They can form athermally during quenching from the β-phase field or isothermally during subsequent aging and often have a substantial influence on mechanical properties, typically increasing strength and elastic modulus while reducing ductility [13,14]. In such titanium alloys, however, the ω phase occurs as nanoscale precipitates embedded in a β-phase matrix. Consequently, experimental measurements of its intrinsic physical properties have been limited. One notable example is the determination of the complete elastic stiffness components of ω-Ti from measurements on a polycrystalline specimen prepared by high-pressure torsion [15].

More recently, static high-pressure synthesis techniques have enabled the recovery of millimeter-sized bulk polycrystalline specimens consisting of single-phase ω-Ti [16-18]. Measurements on such specimens have demonstrated high hardness and elastic modulus [16,17], as well as high compressive strength [17]. Subsequent tensile testing revealed stress-induced ω-to-α transformation during plastic deformation and the resulting transformation-induced plasticity [18]. These advances make it possible to investigate the intrinsic physical properties of bulk ω-Ti under ambient conditions. The volumetric thermal expansivity of ω-Ti has previously been derived from X-ray diffraction measurements performed under high-pressure and high-temperature conditions within its stability field, at approximately 8 GPa [19]. However, neither the axial thermal expansion nor the low-temperature volumetric expansion of ω-Ti has been determined experimentally at ambient pressure.

In the present study, we investigate the thermal expansion of single-phase bulk polycrystalline α- and ω-Ti by X-ray diffraction measurements between approximately 5 and 300 K. The two phases are examined under nearly identical experimental conditions, allowing their thermal expansion behaviors to be compared directly. The temperature dependences of the $a$ and $c$ lattice parameters, unit-cell volume, and $c/a$ ratio are determined, and the axial and volumetric thermal-expansion coefficients are evaluated. The volume data are analyzed using a simple Debye-Grüneisen model. First-principles calculations are also performed for both the α and ω phases to determine their axial and volumetric thermal expansion coefficients. The purpose of this study is to reexamine the low-temperature $c$-axis negative thermal expansion of α-Ti and to experimentally determine the low-temperature thermal expansion behavior of ω-Ti at ambient pressure.

## 2. Experimental and computational details

A bulk polycrystalline ω-Ti specimen was synthesized using a 2000-ton belt-type high-pressure apparatus installed at the National Institute for Materials Science (NIMS), Japan [20]. A commercially available CP-Ti Grade 2 rod, 5 mm in diameter and 11 mm in length, was used as the starting material. The starting material was placed in an hBN sample capsule. The sample was first compressed to 7.7 GPa at room temperature and then heated to 500 °C at a rate of 50 °C/min. After being held at this temperature for 30 min, the sample was quenched to room temperature while the applied load was maintained. The pressure was subsequently released over approximately 1.5 h. The recovered cylindrical ω-Ti specimen was approximately 6 mm in diameter and 7.5 mm in length. A rectangular plate with a thickness of 1.2 mm was cut parallel to the cylinder axis by wire electrical discharge machining (wire EDM) and subsequently chemically polished to remove the machining-damaged surface layer using a commercially available solution containing a mixture of nitric and hydrofluoric acids (SCLEAN S-22, Sasaki Chemical Co., Ltd., Kyoto, Japan) [17]. For comparison, an α-Ti specimen was prepared from the same CP-Ti Grade 2 material in the form of a cylinder 8 mm in diameter and 10 mm in length. A 1.2-mm-thick plate was cut parallel to the cylinder axis by wire EDM and chemically polished in the same manner for the X-ray diffraction (XRD) measurements.

Low-temperature XRD measurements were performed using a SmartLab 9-kW diffractometer equipped with a cryostat (Rigaku Corporation, Tokyo, Japan). X-rays were generated using a rotating Cu anode, and a Johansson monochromator was installed on the incident-beam side to obtain monochromatic Cu $K\alpha_1$ radiation. Each specimen was mounted on a single-crystal Cu zero-diffraction plate and fixed using a small amount of Apiezon N grease. After mounting, the sample chamber was evacuated to below $5\times10^{-5}$ Pa, and the specimen was cooled to 5 K at a rate of 30 K/min. Before the XRD measurements were initiated, the specimen was held at 5 K for at least 6 h to ensure thermal stabilization. The first diffraction pattern was collected at 5 K over a $2\theta$ range of 5°-130°, with a scan rate of 2°/min and a step size of 0.02°. After completion of the measurement, the specimen was heated to 10 K at a rate of 2 K/min and held at that temperature for 30 min. A sample z-scan was then performed to position the specimen surface at the diffraction center, after which an XRD pattern was collected under the same conditions as those used at 5 K. This procedure was subsequently repeated at 10 K intervals during heating up to 300 K. In total, 31 diffraction patterns were collected between 5 and 300 K. The XRD patterns were analyzed using PDIndexer [21] to determine the lattice parameters and unit-cell volumes.

First-principles calculations were performed for α-Ti and ω-Ti using density functional theory within the projector augmented wave method [22] and the generalized gradient approximation in the Perdew-Burke-Ernzerhof form [23], as implemented in the VASP code [24-26]. The elastic stiffness constants listed in Table 5 were calculated using a plane-wave cutoff energy of 500 eV and first-order Methfessel–Paxton smearing [27] with a width of 0.2 eV. The Brillouin zones were sampled with 34 × 34 × 18 and 27 × 27 × 38 k-point meshes for α-Ti and ω-Ti, respectively. These values do not include phonon contributions. To evaluate the axial thermal expansion of the two hexagonal phases, the Helmholtz free energy, i.e., the sum of the static internal energy and the electronic and vibrational free energies, was computed as a function of the lattice parameters $a$ and $c$ on a 5×5 grid of unit cells. At each temperature, the free energies at the 25 grid points were fitted with a polynomial of total degree three in a and c, and the equilibrium values of a and c were taken as the minimum of the fitted surface. The resulting temperature-dependent lattice parameters were fitted to an Einstein model [28-30] consisting of a linear combination of three Einstein terms, from which the axial thermal expansion coefficients were derived. Temperature-dependent force constants were obtained using the stochastic self-consistent harmonic approximation [31,32] as implemented in the phonopy code [33,34], together with the force-constant calculator implemented in the symfc code [35] and the polynomial machine-learning potentials (MLPs) implemented in the pypolymlp code [36]. The polynomial MLPs were trained using the same set of parameters as in Ref. [37] on a training dataset of 180 thermally displaced supercells computed from first-principles calculations. Details of the computational methods and convergence tests are given elsewhere [38].

## 3. Results and discussion

Fig. 1 shows the XRD patterns of α-Ti and ω-Ti collected at the lowest measurement temperatures and at 150 and 300 K. No phase transformation or observable change in the relative intensities of diffracted peaks was observed for either sample over the entire temperature range. For α-Ti, the relative intensities differ from those of the calculated powder pattern, most notably in the enhanced intensity of the 002 reflection. This change is likely associated with the microstructure developed during processing of the α-Ti rod, such as forging or extrusion, in which the *c* axes are preferentially oriented perpendicular to the rod axis. In contrast, the relative intensities of the bulk ω-Ti sample are close to those of the calculated powder pattern, indicating that the preferred orientation present in the starting α-Ti was lost during the α-ω transformation under high-pressure and high-temperature conditions. A similar disappearance of the preferred orientation through nucleation and growth of the ω phase has previously been demonstrated by two-dimensional XRD measurements [17]. To assess the phase purity of the ω-Ti specimen, Rietveld refinements were performed on the XRD patterns collected at 5.7 and 300 K. Both patterns were satisfactorily fitted using a single-phase ω-Ti model. An additional refinement of the 300 K pattern using a two-phase ω-Ti + α-Ti model yielded phase fractions of 100.0(1) wt% for ω-Ti and 0.0(1) wt% for α-Ti, without improving the fit. These results indicate that the bulk polycrystalline specimen examined in this study consists of single-phase ω-Ti within the detection limits of the present low-temperature XRD measurements. The results of the Rietveld refinements are shown in Fig. S-1 of the Supplemental Material. The lattice parameters of the hexagonal lattices, *a* and *c*, were determined for all diffraction patterns, from which the unit-cell volumes and *c*/*a* ratios were calculated. The resulting values are summarized in Tables 1 and 2 for α-Ti and ω-Ti, respectively.

Figs. 2(a) and 2(b) show the unit-cell volumes, $V(T)$, of α-Ti and ω-Ti, respectively, as functions of temperature. The temperature dependence of the unit-cell volume was analyzed using the first-order Grüneisen approximation for the zero-pressure equation of state, in which thermal expansion is treated as an elastic strain [39,40]:

$$V(T) = V_0 + \frac{\gamma}{K_0} U(T) \qquad (1)$$

where $V_0$ is the hypothetical volume at $T = 0$ K, γ is the effective Grüneisen parameter, assumed to be independent of temperature and pressure, and $K_0$ is the isothermal bulk modulus at the ambient pressure. The internal energy, $U(T)$, associated with lattice vibrations was described using the Debye model:

$$U(T) = 9Nk_BT(\frac{T}{\theta_D})^3 \int_0^{\theta_D/T} \frac{x^3}{e^x-1}dx \qquad (2)$$

where $N$ is the number of atoms in the unit cell ($N$ =2 for α-Ti and $N$ =3 for ω-Ti), $k_B$ is the Boltzmann constant, and $\theta_D$ is the Debye temperature. The values of $K_0$ were fixed at 114 and 107 GPa for α-Ti and ω-Ti, respectively [19], and were assumed to be temperature independent between 0 and 300 K. The experimental data were fitted using Eqs. (1) and (2), with $V_0$, γ, and $\theta_D$ as fitting parameters. The resulting model curves are shown as gray solid lines in Figs. 2(a) and 2(b). Table 3 summarizes the Debye temperatures and effective Grüneisen parameters determined for α-Ti and ω-Ti, together with values reported in previous experimental and theoretical studies. For α-Ti, the values obtained in the present study are in excellent agreement with those determined in previous experimental studies [41-43]. To our knowledge, the present study provides the first experimental determination of these parameters for ω-Ti. The experimentally determined Debye temperature of ω-Ti lies between the theoretical estimates derived from calculations based on elasticity and heat capacity [44]. Fig. 2(c) compares the relative volume changes, $(V_T - V_{300})/V_{300}$ of α-Ti and ω-Ti as functions of temperature, where $V_T$ and $V_{300}$ are unit-cell volumes at temperature $T$ and at 300 K, respectively. The two phases exhibit very similar relative volume changes over the entire temperature range within the experimental uncertainties.

Fig. 3 shows the temperature dependences of the lattice parameters *a* and *c* of α-Ti and ω-Ti. The *a*-axis data of α-Ti and both the *a*- and *c*-axis data of ω-Ti were fitted using expressions analogous to Eqs. (1) and (2). For these fits, the Debye temperatures were fixed at the values determined from the unit-cell-volume analyses and only the coefficient relating the lattice parameter to the internal energy was treated as a fitting parameter. The resulting curves are shown as blue solid lines for the *a*-axis data and as a red solid line for the *c*-axis data of ω-Ti. In contrast, the *c*-axis data of α-Ti exhibit a distinct nonmonotonic temperature dependence at low temperatures, indicating negative thermal expansion. The *c*-axis data were therefore fitted empirically using third- and fifth-order polynomial functions [45]. These polynomial functions were used solely to reproduce the observed temperature dependence and were not intended to represent a physical model. The fifth-order polynomial reproduced the low-temperature behavior more accurately, whereas the third-order polynomial provided a better representation at higher temperatures. Accordingly, the fifth-order polynomial was used from 5.2 to 120 K and the third-order polynomial above 120 K. The resulting curve is shown as the red solid line in Fig. 3(c). The lattice parameters reported by Hope and Kodess [8] at 10 and 290 K, also shown in Figs. 3(a) and 3(c), are in close agreement with the present values. As shown in the inset, the present *c*-axis parameter initially decreases with increasing temperature from the lowest-temperature value and reaches a shallow minimum before increasing at higher temperatures.

Fig. 4 shows the temperature dependences of the *c*/*a* ratios of α-Ti and ω-Ti. For α-Ti, *c*/*a* decreases continuously with increasing temperature, reflecting the pronounced anisotropic thermal expansion of the *a* and *c* axes. In contrast, the *c*/*a* ratio of ω-Ti remains consistent with the ideal bcc-derived value of $\sqrt{3/8}$ = 0.61237 [46,47] within the experimental uncertainties over the entire temperature range, although a slight overall decrease with increasing temperature is discernible. The contrasting behaviors of the two phases are also consistent with their lattice responses under compression [48]. Errandonea et al. [48] showed that the *c*/*a* ratio of α-Ti increases markedly with increasing pressure, whereas that of ω-Ti changes only slightly even under compression up to 16 GPa, from 0.609 at ambient pressure to 0.613 at 16 GPa. Taken together, the temperature and pressure dependences of *c*/*a* demonstrate a pronounced anisotropic lattice response in α-Ti, whereas ω-Ti exhibits a nearly isotropic lattice response, with *c*/*a* remaining remarkably stable at approximately 0.61 under both thermal contraction and mechanical compression.

Fig. 5 summarizes the volumetric and linear thermal expansion coefficients (TECs) of α-Ti and ω-Ti as functions of temperature. The volumetric TECs were calculated from the temperature derivatives of the fitted unit-cell volume functions, *V*(T), shown in Figs. 2(a) and 2(b), according to

$$\alpha_V(T) = \frac{1}{V(T)}\frac{dV(T)}{dT} \ . \qquad (3)$$

The linear TECs along the *a* axes were calculated from the temperature derivatives of the fitted lattice-parameter functions, *a*(T), shown in Figs. 3(a) and 3(b), according to

$$\alpha_a(T) = \frac{1}{a(T)}\frac{da(T)}{dT} \ . \qquad (4)$$

The linear TECs along the *c* axes were calculated in the same manner using the fitted *c*(T) functions shown in Figs. 3(c) and 3(d). The TECs were evaluated at temperatures corresponding to the XRD measurements. Figs. 5(a) and 5(b) show the volumetric TECs of α-Ti and ω-Ti, respectively. The two phases exhibit very similar temperature dependences over the entire temperature range. In contrast, their linear TECs show markedly different behaviors, as shown in Figs. 5(c) and 5(d). For α-Ti, the *c*-axis TEC is substantially smaller than the *a*- axis TEC and becomes negative at low temperatures. The *c*-axis TEC reaches a minimum value of $-0.71 \times 10^{-6}$ $K^{-1}$ at 20 K and remains negative below approximately 50 K. For ω-Ti, the *a*- and *c*-axis TECs are very similar over the entire temperature range, indicating nearly isotropic thermal expansion, although the *a*-axis TEC is slightly larger than the *c*-axis TEC. Thus, despite the markedly different axial thermal-expansion behaviors of α-Ti and ω-Ti, their volumetric thermal-expansion behaviors are very similar.

Fig. 6 shows the temperature dependences of the TECs of α-Ti and ω-Ti determined experimentally and theoretically in the present study. The corresponding numerical data are provided in Tables S-1 to S-4 of the Supplemental Material. For α-Ti, several experimental [3,5,49,50] and theoretical studies [6,7] have previously reported TECs, and their results are also included in Fig. 6(a) for comparison. For ω-Ti, the theoretically predicted volumetric TEC reported by Argaman et al. [44] is included in Fig. 6(b). For the volumetric TECs of both phases, the experimental and theoretical results, including those from previous studies, are generally in good agreement; however, the reported axial TECs, particularly those of α-Ti, show considerably larger variations among both experimental and theoretical studies. To our knowledge, Nizhankovskii et al. [5] provided the only previous experimental measurements of both the *a*- and *c*-axis linear thermal expansion coefficients of α-Ti at low temperatures. They reported pronounced negative thermal expansion along the *c* axis, with the *c*-axis TEC reaching approximately $-9.5\times10^{-6}$ $K^{-1}$ at around 150 K. They also reported a sharp increase in the *c*-axis TEC from negative to positive values between 150 and 180 K. In contrast, the magnitudes of the negative *c*-axis TECs obtained both experimentally and theoretically in the present study are substantially smaller than the magnitude reported by Nizhankovskii et al. [5]. The present experimental *c*-axis TEC reaches a minimum of $-0.71\times10^{-6}$ $K^{-1}$ at 20 K, while the present theoretical calculation also shows negative *c*-axis thermal expansion, with a minimum of $-3.04\times10^{-6}$ $K^{-1}$ at 50 K and negative values extending to approximately 105 K. Although the origin of the discrepancy with Nizhankovskii et al. [5] remains unclear, possible effects of internal strain in the present textured polycrystalline α-Ti specimen cannot be excluded. High-precision laser-interferometric measurements of the length changes along the *a* and *c* axes of single-crystal α-Ti, as demonstrated for single-crystal Si [51], would provide an independent test of the magnitude and temperature range of the *c*-axis negative thermal expansion. For ω-Ti, the volumetric TECs obtained experimentally and theoretically in the present study are in excellent agreement, with values of $25.09 \times 10^{-6}$ $K^{-1}$ and $25.02 \times 10^{-6}$ $K^{-1}$ at 300 K, respectively. For comparison, Zhang et al. [19] reported a volumetric TEC of approximately $14.8 \times 10^{-6}$ $K^{-1}$ at 8.1 GPa and 300 K, derived from their high-pressure thermal equation of state. This value is lower than the present ambient-pressure value, consistent with the expected suppression of thermal expansion under compression, partly due to the increase in the bulk modulus with increasing pressure. Regarding the axial TECs, both the experimental and theoretical results show that the *a*-axis TEC is larger than the *c*-axis TEC. However, while the experimental *a*- and *c*-axis TECs are nearly identical, the theoretical calculations predict a more pronounced anisotropy.

To quantitatively compare the anisotropy of the axial thermal expansion, Table 4 summarizes the *a*- and *c*-axis TECs and their ratios, $\alpha_c/\alpha_a$, at approximately 300 K. For α-Ti, the ratios obtained experimentally and theoretically in the present study are 0.69 and 0.59, respectively. Previous experimental and theoretical studies also reported ratios below 1, with values ranging from 0.13 to 0.65. In contrast, the ratios for ω-Ti are 0.98 (experiment) and 0.75 (theory) in the present study. Although the theoretical calculations predict greater anisotropy than observed experimentally for ω-Ti, both experiment and theory consistently show that the axial thermal expansion of ω-Ti is more isotropic than that of α-Ti.

The axial thermal expansion of hexagonal crystals is influenced by their elastic anisotropy. We therefore calculated the single-crystal elastic constants of α-Ti and ω-Ti. Table 5 compares the calculated elastic constants with previously reported experimental and theoretical values. The present results are in good agreement with previous first-principles calculations. To characterize the elastic anisotropy, we evaluated three dimensionless parameters commonly used for hexagonal crystals [54,55]: $C_{33} / C_{11}$, $A_{pr} = 4C_{44} / (C_{11} + C_{33} - 2C_{13})$, and $C_{44} / C_{66}$. Here, $C_{33} / C_{11}$ compares the longitudinal elastic stiffness constants along the *c* and *a* axes, $A_{pr}$ represents the shear anisotropy factor associated with prismatic planes, and $C_{44} / C_{66}$ is a Zener-type shear anisotropy ratio comparing the shear stiffness involving the *c* axis with that within the basal plane. All three parameters are equal to 1 in the elastically isotropic limit. The calculated $C_{33} / C_{11}$ and $A_{pr}$ indicate greater elastic anisotropy in ω-Ti than in α-Ti, despite the more isotropic axial thermal expansion of ω-Ti. This suggests that the contrasting axial thermal expansion behaviors of the two phases cannot be explained by elastic anisotropy alone and that temperature-dependent thermal driving forces also play an important role.

The contributions of elastic coupling and temperature-dependent thermal driving forces to the axial thermal expansion can be expressed by the Grüneisen relation for hexagonal crystals, $\alpha_c = 2s_{13}\lambda_a + s_{33}\lambda_c$, where $s_{ij}$ are the elastic compliances and $\lambda_a$ and $\lambda_c$ represent the thermal driving terms associated with strains along the *a* and *c* axes, respectively [38,56]. In the harmonic approximation, these thermal driving terms can be expressed as heat-capacity-weighted sums of directional mode Grüneisen parameters. Table 6 summarizes the contributions of the two terms to the calculated *c*-axis TECs of α-Ti and ω-Ti, as reported in Ref. [38]. The contribution associated with the *a* axis is negative for both phases at all temperatures listed in Table 6. At 50 K, the contribution associated with the *c* axis is also negative for α-Ti, resulting in negative *c*-axis thermal expansion. In contrast, for ω-Ti, the positive contribution associated with the *c* axis exceeds the negative contribution associated with the *a* axis, resulting in positive *c*-axis thermal expansion. At 100 K, the two contributions nearly cancel for α-Ti, making the calculated *c*-axis TEC particularly sensitive to small changes in the computational conditions. These results demonstrate that the contrasting axial thermal expansion behaviors of α-Ti and ω-Ti depend on the balance between elastic coupling and temperature-dependent thermal driving forces. A detailed analysis of this behavior and its computational sensitivity is provided in Ref. [38].

## 4. Conclusions

In conclusion, we investigated the low-temperature thermal expansion of α-Ti and ω-Ti by X-ray diffraction together with first-principles calculations. Although the two phases exhibit very similar volumetric thermal expansion, their axial thermal-expansion behaviors are markedly different. α-Ti shows pronounced anisotropy, including negative thermal expansion along the *c* axis at low temperatures, whereas ω-Ti exhibits nearly isotropic thermal expansion with a $c/a$ ratio that remains close to the ideal bcc-derived value over the entire temperature range. The first-principles calculations reproduce the volumetric thermal expansion of both phases and capture the overall trends of their axial responses. The calculated elastic constants indicate substantial elastic anisotropy in ω-Ti despite its nearly isotropic thermal expansion. The anisotropic thermal expansion of α-Ti and the nearly isotropic thermal expansion of ω-Ti can be understood in terms of the balance between elastic coupling and temperature-dependent thermal driving forces.

## CRediT authorship contribution statement

**Norimasa Nishiyama:** Conceptualization, Methodology, Investigation, Formal analysis, Visualization, Writing – original draft, Writing – review & editing, Supervision, Project administration. **Atsushi Togo:** Conceptualization, Methodology, Investigation, Formal analysis, Writing – original draft, Writing – review & editing. **Takuya Sasaki:** Methodology, Investigation, Formal analysis, Visualization, Writing – review & editing. **Yoshitaka Matsushita:** Methodology, Investigation, Data curation, Writing – review & editing.

## Declaration of competing interest

The authors declare that they have no known competing financial interests or personal relationships that could have appeared to influence the work reported in this paper.

**Acknowledgements**

We thank K. Tsuzaki, M. Wakeda, T. Ohmura, M. Matsushita, and T. Taniguchi for discussion. We also thank T. Ikeda, M. Nakayama, M. Iida, S. Okamoto, and K. Matsuda for technical and administrative assistance. This work was supported by the World Premier International Research Center Initiative (WPI). This work was partly supported by JSPS KAKENHI Grant Numbers JP24K08021, JP24H00190, JP25H01246, and JP25H01252 to A.T. The theoretical calculations in this study were performed on the Numerical Materials Simulator at NIMS. The XRD measurements in this study were supported by "Advanced Research Infrastructure for Materials and Nanotechnology in Japan (ARIM)" of the Ministry of Education, Culture, Sports, Science and Technology (MEXT). Proposal Numbers JPMXP1225NM5488 and JPMXP1226NM5361.

**Data availability**

Data will be made available on request.

**References**

[1] C.J. McHargue, J.P. Hammond, Deformation mechanisms in titanium at elevated temperatures, Acta Metall. 1 (1953) 700–705.

[2] J. Spreadborough, J.W. Christian, The measurement of the lattice expansions and Debye temperatures of titanium and silver by X-ray methods, Proc. Phys. Soc. 74 (1959) 609–615.

[3] R.R. Pawar, V.T. Deshpande, The anisotropy of the thermal expansion of α-titanium, Acta Crystallogr. A 24 (1968) 316–317.

[4] P.I. Mal'ko, D.S. Arensburger, V.S. Pugin, V.F. Nemchenko, S.N. L'vov, Thermal and

electrical properties of porous titanium, Powder Metall. Met. Ceram. 9 (1970) 642–644.

[5] V.I. Nizhankovskii, M.I. Katsnelson, G.V. Peschanskikh, A.V. Trefilov, Anisotropy of the thermal expansion of titanium due to proximity to an electronic topological transition, JETP Lett. 59 (1994) 733–737.

[6] P. Souvatzis, O. Eriksson, M.I. Katsnelson, Anomalous thermal expansion in α-titanium, Phys. Rev. Lett. 99 (2007) 015901.

[7] U. Argaman, E. Eidelstein, O. Levy, G. Makov, Ab initio study of the phononic origin of negative thermal expansion, Phys. Rev. B 94 (2016) 174305.

[8] H. Hope, B. Kodess, Cell dimensions of titanium from 10 K to 290 K, Acta Crystallogr. A 72 (2016) s263.

[9] J.C. Jamieson, Crystal structures of titanium, zirconium, and hafnium at high pressures, Science 140 (1963) 72–73.

[10] A. Dewaele, V. Stutzmann, J. Bouchet, F. Bottin, F. Occelli, M. Mezouar, High-pressure-temperature phase diagram and equation of state of titanium, Phys. Rev. B 91 (2015) 134108.

[11] S. Ankem, C.A. Greene, Recent developments in microstructure/property relationships of beta titanium alloys, Mater. Sci. Eng. A 263 (1999) 127–131.

[12] M.T. Mohammed, Z.A. Khan, A.N. Siddiquee, Beta titanium alloys: The lowest elastic modulus for biomedical applications: A review, Int. J. Chem. Nucl. Metall. Mater. Eng. 8 (2014) 726–731.

[13] B.S. Hickman, The formation of omega phase in titanium and zirconium alloys: A review, J. Mater. Sci. 4 (1969) 554–563.

[14] J.C. Williams, B.S. Hickman, H.L. Marcus, The effect of omega phase on the mechanical properties of titanium alloys, Metall. Trans. 2 (1971) 1913–1919.

[15] M. Tane, Y. Okuda, Y. Todaka, H. Ogi, A. Nagakubo, Elastic properties of single-crystalline ω phase in titanium, Acta Mater. 61 (2013) 7543–7554.

[16] S. Li, Q. Yuan, J. Zhang, Y. Li, D. He, Towards pure: The single-phase bulk omega

titanium and modulation on its elastic properties under biaxial strains, J. Alloys Compd. 906 (2022) 164312.

[17] T. Sawahata, N. Nishiyama, M. Arita, Y. Kawabata, M. Matsushita, K. Ohara, Y. Higo, F. Wakai, Z. Horita, High compressive strength of bulk polycrystalline ω phase in pure titanium, Mater. Trans. 66 (2025) 616–621.

[18] N. Nishiyama, Y. Tange, T. Sawahata, M. Matsushita, K. Tokuda, K. Tominaga, T. Sekiya, K. Kuramochi, K. Sasaki, F. Wakai, Z. Horita, Y. Kobayashi, A. Fujimura, High tensile strength and transformation-induced plasticity in bulk polycrystalline omega titanium, Sci. Rep. 16 (2026) 5395.

[19] J. Zhang, Y. Zhao, R.S. Hixson, G.T. Gray III, L. Wang, W. Utsumi, H. Saito, T. Hattori, Thermal equations of state for titanium obtained by high pressure-temperature diffraction studies, Phys. Rev. B 78 (2008) 054119.

[20] T. Taniguchi, M. Akaishi, S. Yamaoka, Sintering of cubic boron nitride without additives at 7.7 GPa and above 2000 °C, J. Mater. Res. 14 (1999) 162–169.

[21] Y. Seto, D. Nishio-Hamane, T. Nagai, N. Sata, Development of a software suite on X-ray diffraction experiments, Rev. High Pressure Sci. Technol. 20 (2010) 269–276.

[22] P.E. Blöchl, Projector augmented-wave method, Phys. Rev. B 50 (1994) 17953–17979.

[23] J.P. Perdew, K. Burke, M. Ernzerhof, Generalized gradient approximation made simple, Phys. Rev. Lett. 77 (1996) 3865–3868.

[24] G. Kresse, D. Joubert, From ultrasoft pseudopotentials to the projector augmented-wave method, Phys. Rev. B 59 (1999) 1758–1775.

[25] G. Kresse, J. Furthmüller, Efficiency of ab-initio total energy calculations for metals and semiconductors using a plane-wave basis set, Comput. Mater. Sci. 6 (1996) 15–50.

[26] G. Kresse, Ab initio molecular dynamics for liquid metals, J. Non-Cryst. Solids 192–193 (1995) 222–229.

[27] M. Methfessel, A.T. Paxton, High-precision sampling for Brillouin-zone integration in

metals, Phys. Rev. B 40 (1989) 3616–3621.

[28] T. Chatterji, T.C. Hansen, Magnetoelastic effects in Jahn-Teller distorted $CrF_2$ and $CuF_2$ studied by neutron powder diffraction, J. Phys.: Condens. Matter 23 (2011) 276007.

[29] R.J. Angel, M. Alvaro, J. Gonzalez-Platas, EosFit7c and a Fortran module (library) for equation of state calculations, Z. Kristallogr. Cryst. Mater. 229 (2014) 405–419.

[30] D.C. Wallace, Thermodynamics of Crystals, John Wiley & Sons, New York, 1972.

[31] I. Errea, M. Calandra, F. Mauri, First-principles theory of anharmonicity and the inverse isotope effect in superconducting palladium-hydride compounds, Phys. Rev. Lett. 111 (2013) 177002.

[32] A. van Roekeghem, J. Carrete, N. Mingo, Quantum self-consistent ab-initio lattice dynamics, Comput. Phys. Commun. 263 (2021) 107945.

[33] A. Togo, L. Chaput, T. Tadano, I. Tanaka, Implementation strategies in phonopy and phono3py, J. Phys.: Condens. Matter 35 (2023) 353001.

[34] A. Togo, First-principles phonon calculations with Phonopy and Phono3py, J. Phys. Soc. Jpn. 92 (2023) 012001.

[35] A. Seko, A. Togo, Projector-based efficient estimation of force constants, Phys. Rev. B 110 (2024) 214302.

[36] A. Seko, Tutorial: Systematic development of polynomial machine learning potentials for elemental and alloy systems, J. Appl. Phys. 133 (2023) 011101.

[37] A. Togo, A. Seko, On-the-fly training of polynomial machine learning potentials in computing lattice thermal conductivity, J. Chem. Phys. 160 (2024) 211001.

[38] A. Togo, Axial thermal expansion from free-energy minimization with temperature-dependent force constants in phonopy: a technical report, with α- and ω-Ti as the worked example, arXiv:2609.24336 (2026).

[39] D.M. Trots, A. Kurnosov, T. Boffa Ballaran, D.J. Frost, High-temperature structural behaviors of anhydrous wadsleyite and forsterite, Am. Mineral. 97 (2012) 1582–1590.

[40] K.S. Knight, A.S. Gibbs, C.L. Bull, A.V. Powell, N.P. Funnell, C.J. Ridley, Low-intermediate-temperature, high-pressure thermoelastic and crystallographic properties of thermoelectric clausthalite (PbSe-I), Mater. Adv. 3 (2022) 2077–2088.

[41] E.S. Fisher, C.J. Renken, Single-crystal elastic moduli and the hcp → bcc transformation in Ti, Zr, and Hf, Phys. Rev. 135 (1964) A482–A494.

[42] G.D. Kneip Jr., J.O. Betterton Jr., J.O. Scarbrough, Low-temperature specific heats of titanium, zirconium, and hafnium, Phys. Rev. 130 (1963) 1687–1692.

[43] K.A. Gschneidner Jr., Physical properties and interrelationships of metallic and semimetallic elements, Solid State Phys. 16 (1964) 275–426.

[44] U. Argaman, E. Eidelstein, O. Levy, G. Makov, Thermodynamic properties of titanium from ab initio calculations, Mater. Res. Express 2 (2015) 016505.

[45] K. Röttger, A. Endriss, J. Ihringer, S. Doyle, W.F. Kuhs, Lattice constants and thermal expansion of $H_2O$ and $D_2O$ Ice Ih between 10 and 265 K. Addendum, Acta Crystallogr. B 68 (2012) 91.

[46] S.G. MacLeod, B.E. Tegner, H. Cynn, W.J. Evans, J.E. Proctor, M.I. McMahon, G.J. Ackland, Experimental and theoretical study of Ti-6Al-4V to 220 GPa, Phys. Rev. B 85 (2012) 224202.

[47] T. Li, D. Kent, G. Sha, L.T. Stephenson, A.V. Ceguerra, S.P. Ringer, M.S. Dargusch, J.M. Cairney, New insights into the phase transformations to isothermal ω and ω-assisted α in near β-Ti alloys, Acta Mater. 106 (2016) 353–366.

[48] D. Errandonea, Y. Meng, M. Somayazulu, D. Häusermann, Pressure-induced α→ω transition in titanium metal: A systematic study of effects of uniaxial stress, Physica B 355 (2005) 116–125.

[49] Y.S. Touloukian, R.K. Kirby, R.E. Taylor, P.D. Desai, Thermophysical Properties of Matter—The TPRC Data Series, Vol. 12: Thermal Expansion—Metallic Elements and Alloys, IFI/Plenum, New York, 1975.

[50] J.A. Cowan, A.T. Pawlowicz, G.K. White, Thermal expansion of polycrystalline titanium and zirconium, Cryogenics 8 (1968) 155–157.

[51] T. Middelmann, A. Walkov, G. Bartl, R. Schödel, Thermal expansion coefficient of single-crystal silicon from 7 K to 293 K, Phys. Rev. B 92 (2015) 174113.

[52] D.R. Trinkle, M.D. Jones, R.G. Hennig, S.P. Rudin, R.C. Albers, J.W. Wilkins, Empirical tight-binding model for titanium phase transformations, Phys. Rev. B 73 (2006) 094123.

[53] C.S. Allen, A.P. Bartók, Multi-phase dataset for bulk Ti and the Ti-6Al-4V alloy, Sci. Data 12 (2025) 1925. DOI: 10.1038/s41597-025-05302-3

[54] K. Kishida, M. Fujiwara, H. Adachi, K. Tanaka, H. Inui, Plastic deformation of single crystals of $Ti_5Si_3$ with the hexagonal $D8_8$ structure, Acta Mater. 58 (2010) 846–857.

[55] X.W. Xu, L. Hu, X. Yu, Z.M. Lu, Y. Fan, Y.X. Li, C.C. Tang, Ab initio study of the electronic structure and elastic properties of $Al_5C_3N$, Chin. Phys. B 20 (2011) 126201.

[56] M. Xu, Q. Gao, Y. Qiao, K. Zhao, A. Sanson, Q. Sun, C. Wang, A. Venier, F. d'Acapito, M. Tucker, Y. Song, C. Zhou, X. Xing, J. Chen, Quantitative role of phonons and elasticity in tuning uniaxial negative thermal expansion of $MZr_2$ (M = Fe, Co, and Ni), Phys. Rev. Lett. 135 (2025) 166101.

**Figure captions**

**Fig. 1.** XRD patterns at selected temperatures: (a) α-Ti; (b) ω-Ti. The asterisks indicate the reflections from the Cu sample holder.

**Fig. 2.** Temperature dependence of the unit-cell volumes of (a) α-Ti and (b) ω-Ti. (c) The temperature dependence of the relative unit-cell volume changes of α-Ti and ω-Ti. In (c), the relative unit-cell volume changes are expressed as a percentage.

**Fig. 3.** Temperature dependence of the lattice parameters: (a) *a*-axis of α-Ti; (b) *a*-axis of ω-

Ti; (c) *c*-axis of α-Ti; (d) *c*-axis of ω-Ti. In (a) and (c), data reported in a previous study [8] are also shown for comparison. The inset in (c) shows an expanded view up to 130 K.

**Fig. 4.** Temperature dependence of the *c*/*a* ratios of (a) α-Ti and (b) ω-Ti. The horizontal line in (b) indicates the ideal bcc-derived value, $\sqrt{3/8}$ = 0.61237.

**Fig. 5.** Temperature dependence of thermal expansion coefficients (TECs): (a) volumetric TEC of α-Ti; (b) volumetric TEC of ω-Ti; (c) linear TECs of α-Ti; (d) linear TECs of ω-Ti.

**Fig. 6.** Temperature dependence of the thermal expansion coefficients of (a) α-Ti and (b) ω-Ti determined experimentally and theoretically in the present study. Previously reported experimental and theoretical values are also shown for comparison (see text).

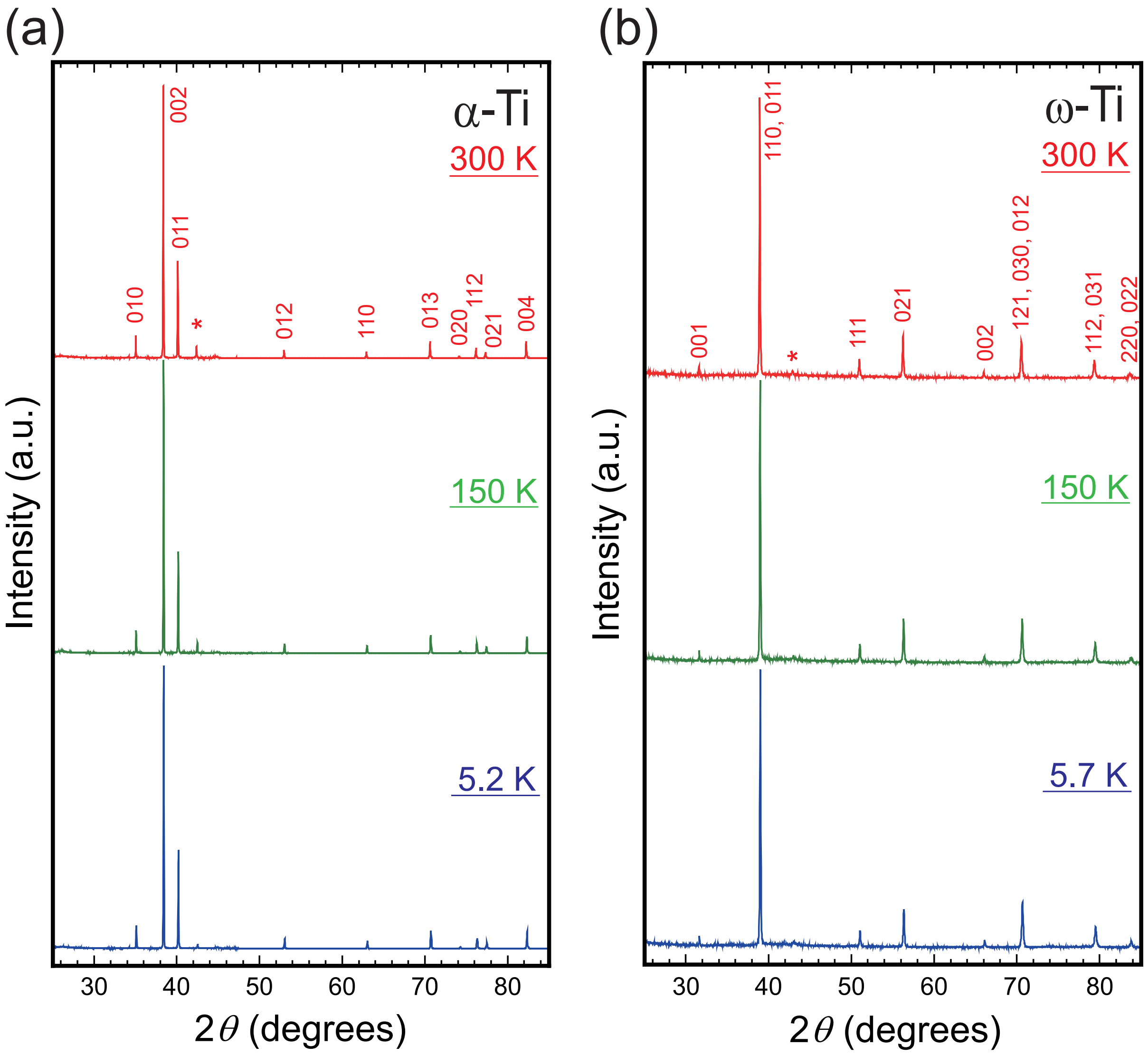

**Fig. 1.** XRD patterns at selected temperatures: (a) α-Ti; (b) ω-Ti. The asterisks indicate the reflections from the Cu sample holder.

Figure 1

**Table 1**

Unit cell parameters, unit cell volumes, and $c/a$ ratios of α-Ti.

| Temperature (K) | $a$ (Å) | $c$ (Å) | $V$ (Å$^3$) | $c/a$ |
|---|---|---|---|---|
| 5.2 | 2.9464(2) | 4.6798(4) | 35.185(13) | 1.5883(2) |
| 9.9 | 2.9465(2) | 4.6798(4) | 35.185(13) | 1.5883(2) |
| 20.0 | 2.9464(2) | 4.6798(4) | 35.184(13) | 1.5883(2) |
| 29.9 | 2.9464(2) | 4.6797(4) | 35.184(14) | 1.5883(2) |
| 40.0 | 2.9464(2) | 4.6797(4) | 35.184(14) | 1.5883(2) |
| 50.0 | 2.9465(2) | 4.6797(4) | 35.186(13) | 1.5882(2) |
| 59.9 | 2.9466(2) | 4.6797(4) | 35.187(14) | 1.5882(2) |
| 70.0 | 2.9467(2) | 4.6797(4) | 35.190(14) | 1.5881(2) |
| 79.9 | 2.9467(2) | 4.6797(4) | 35.191(14) | 1.5881(2) |
| 91.0 | 2.9469(2) | 4.6798(4) | 35.195(14) | 1.5880(2) |
| 100.0 | 2.9470(2) | 4.6800(4) | 35.200(13) | 1.5880(2) |
| 110.0 | 2.9472(2) | 4.6801(4) | 35.204(14) | 1.5880(2) |
| 120.0 | 2.9473(2) | 4.6802(4) | 35.209(14) | 1.5880(2) |
| 130.0 | 2.9475(2) | 4.6804(4) | 35.216(13) | 1.5879(2) |
| 140.0 | 2.9478(2) | 4.6806(4) | 35.222(14) | 1.5879(2) |
| 150.0 | 2.9479(2) | 4.6808(4) | 35.228(14) | 1.5878(2) |
| 160.0 | 2.9482(2) | 4.6810(4) | 35.236(14) | 1.5877(2) |
| 170.0 | 2.9484(2) | 4.6812(4) | 35.242(14) | 1.5877(2) |
| 180.0 | 2.9486(2) | 4.6815(4) | 35.249(14) | 1.5877(2) |
| 190.0 | 2.9488(2) | 4.6818(4) | 35.257(13) | 1.5877(2) |
| 200.0 | 2.9491(2) | 4.6820(4) | 35.264(14) | 1.5876(2) |
| 210.0 | 2.9493(2) | 4.6823(4) | 35.272(14) | 1.5876(2) |
| 220.0 | 2.9495(2) | 4.6826(4) | 35.280(14) | 1.5876(2) |
| 230.0 | 2.9498(2) | 4.6828(4) | 35.288(14) | 1.5875(2) |
| 240.0 | 2.9500(2) | 4.6832(4) | 35.296(13) | 1.5875(2) |
| 250.0 | 2.9503(2) | 4.6835(4) | 35.304(13) | 1.5875(2) |
| 260.0 | 2.9505(2) | 4.6837(4) | 35.312(14) | 1.5874(2) |
| 270.0 | 2.9508(2) | 4.6840(4) | 35.321(14) | 1.5873(2) |
| 280.0 | 2.9511(2) | 4.6843(4) | 35.330(13) | 1.5873(2) |
| 290.0 | 2.9514(2) | 4.6846(4) | 35.340(13) | 1.5873(2) |
| 300.0 | 2.9516(2) | 4.6850(4) | 35.348(14) | 1.5872(2) |

**Table 2**

Unit cell parameters, unit cell volumes, and $c/a$ ratios of ω-Ti.

| Temperature (K) | $a$ (Å) | $c$ (Å) | $V$ (Å$^3$) | $c/a$ |
|---|---|---|---|---|
| 5.7 | 4.6128(3) | 2.8247(2) | 52.051(15) | 0.61237(6) |
| 10.0 | 4.6125(3) | 2.8245(2) | 52.042(16) | 0.61236(6) |
| 20.0 | 4.6127(3) | 2.8246(2) | 52.047(16) | 0.61236(6) |
| 29.9 | 4.6126(3) | 2.8245(2) | 52.044(16) | 0.61235(6) |
| 39.9 | 4.6128(3) | 2.8246(2) | 52.048(15) | 0.61233(6) |
| 49.9 | 4.6126(3) | 2.8245(2) | 52.043(16) | 0.61235(6) |
| 60.0 | 4.6128(3) | 2.8246(2) | 52.051(17) | 0.61234(7) |
| 69.9 | 4.6130(3) | 2.8248(2) | 52.057(15) | 0.61236(6) |
| 79.9 | 4.6131(3) | 2.8248(2) | 52.062(16) | 0.61235(6) |
| 90.0 | 4.6133(3) | 2.8250(2) | 52.069(15) | 0.61235(6) |
| 100.0 | 4.6134(3) | 2.8251(2) | 52.072(17) | 0.61236(7) |
| 110.0 | 4.6136(3) | 2.8252(2) | 52.080(16) | 0.61236(6) |
| 120.0 | 4.6140(3) | 2.8253(2) | 52.089(16) | 0.61233(6) |
| 130.0 | 4.6142(3) | 2.8255(2) | 52.098(15) | 0.61235(6) |
| 140.0 | 4.6145(3) | 2.8258(2) | 52.109(17) | 0.61236(7) |
| 150.0 | 4.6148(3) | 2.8258(2) | 52.116(16) | 0.61233(6) |
| 160.0 | 4.6150(3) | 2.8260(2) | 52.125(16) | 0.61235(6) |
| 170.0 | 4.6154(3) | 2.8262(2) | 52.138(16) | 0.61236(6) |
| 180.0 | 4.6156(3) | 2.8264(2) | 52.146(17) | 0.61234(7) |
| 190.0 | 4.6161(3) | 2.8266(2) | 52.160(16) | 0.61233(6) |
| 200.0 | 4.6163(3) | 2.8267(2) | 52.167(15) | 0.61234(6) |
| 210.0 | 4.6166(3) | 2.8269(2) | 52.177(16) | 0.61234(6) |
| 220.0 | 4.6170(3) | 2.8271(2) | 52.191(16) | 0.61233(6) |
| 230.0 | 4.6174(3) | 2.8275(2) | 52.208(16) | 0.61236(6) |
| 240.0 | 4.6178(3) | 2.8277(2) | 52.221(16) | 0.61236(6) |
| 250.0 | 4.6181(3) | 2.8278(2) | 52.230(16) | 0.61233(6) |
| 260.0 | 4.6186(3) | 2.8281(2) | 52.245(15) | 0.61232(6) |
| 270.0 | 4.6190(3) | 2.8284(2) | 52.260(15) | 0.61233(6) |
| 280.0 | 4.6193(3) | 2.8286(2) | 52.269(17) | 0.61234(7) |
| 290.0 | 4.6198(3) | 2.8288(2) | 52.285(17) | 0.61233(7) |
| 300.0 | 4.6202(3) | 2.8291(2) | 52.300(16) | 0.61233(6) |

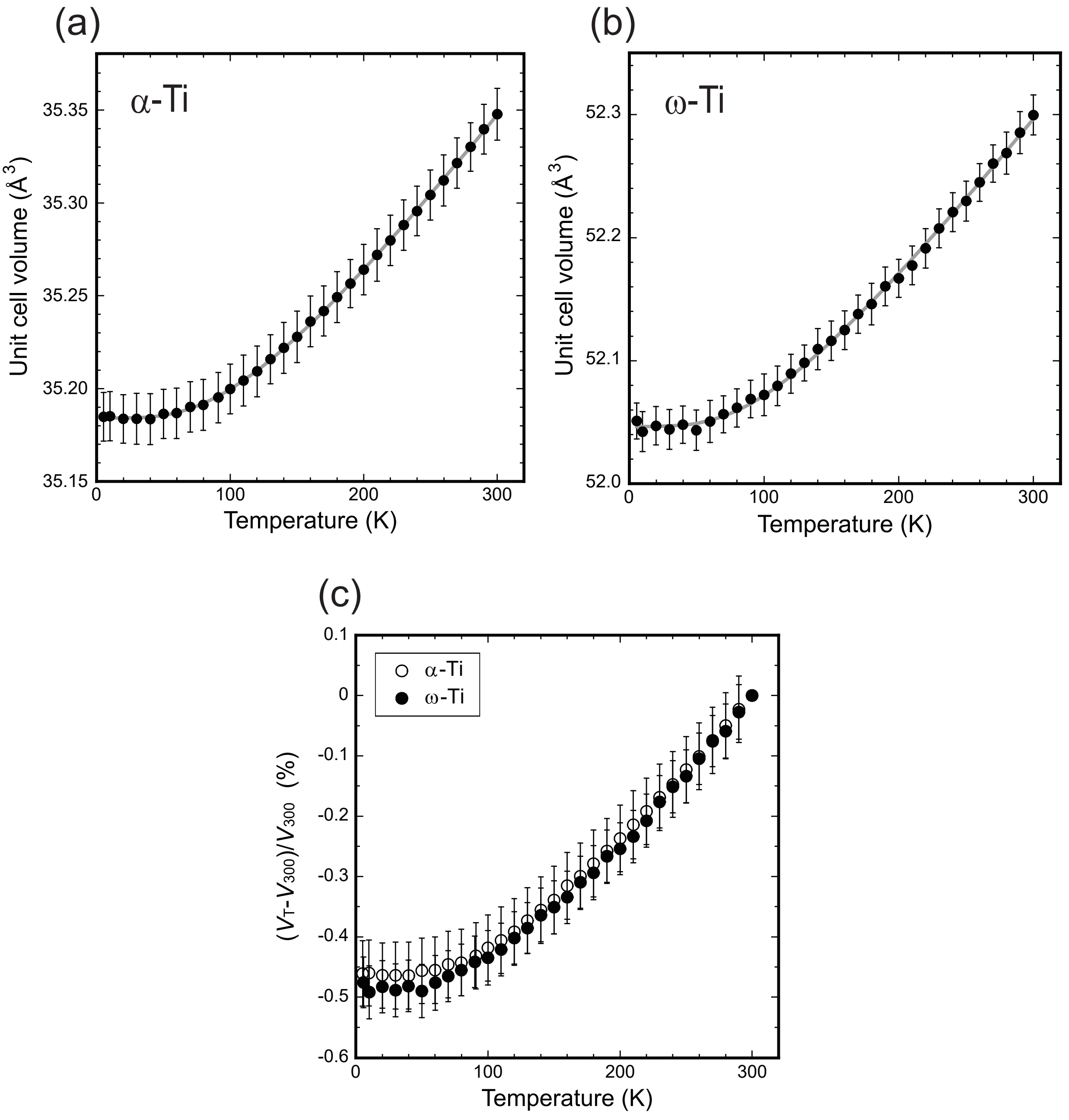

**Fig. 2.** Temperature dependence of the unit-cell volumes of (a) α-Ti and (b) ω-Ti. (c) The temperature dependence of the relative unit-cell volume changes of α-Ti and ω-Ti. In (c), the relative unit-cell volume changes are expressed as a percentage.

Figure 2

**Table 3**

Debye temperatures and effective Grüneisen parameters of α-Ti and ω-Ti determined in the present and previous studies.

| Phase | $\theta_D$ (K) | γ | Remarks |
|---|---|---|---|
| α-Ti | 427(5) | 1.33(4) | *Exp.*, Present study |
| | 426 | - | *Exp.*, Elasticity [41] |
| | 427(5) | - | *Exp.*, Heat capacity [42] |
| | - | 1.33 | *Exp.*, [43] |
| | 403 | - | *Theo.*, Elasticity [44] |
| | 350 | - | *Theo.*, Heat capacity [44] |
| ω-Ti | 409(12) | 1.23(4) | *Exp.*, Present study |
| | 457 | - | *Theo.*, Elasticity [44] |
| | 356 | - | *Theo.*, Heat capacity [44] |

*Exp.*, Experimental study; *Theo.*, Theoretical study

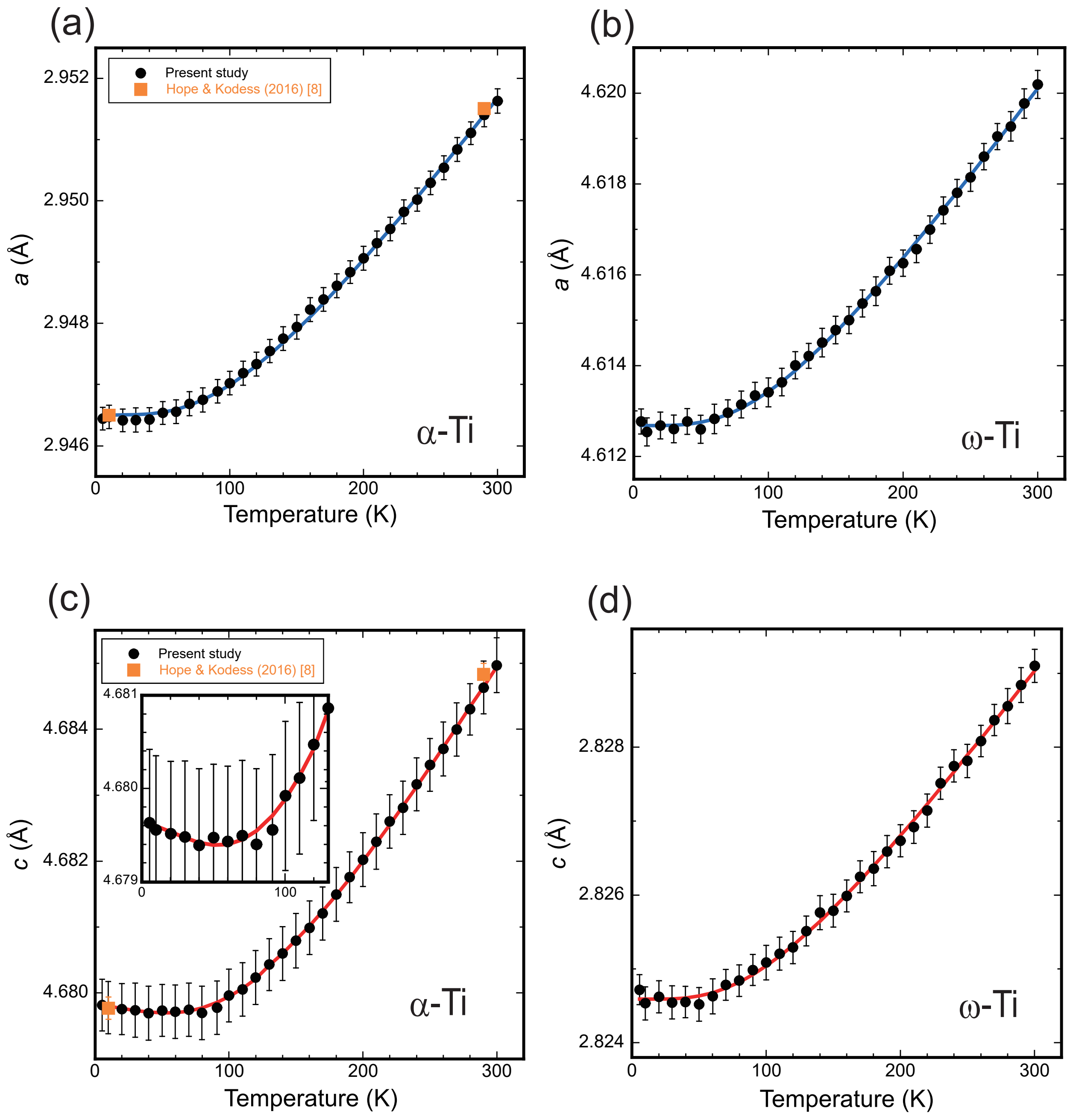

**Fig. 3.** Temperature dependence of the lattice parameters: (a) *a*-axis of α-Ti; (b) *a*-axis of ω-Ti; (c) *c*-axis of α-Ti; (d) *c*-axis of ω-Ti. In (a) and (c), data reported in a previous study [8] are also shown for comparison. The inset in (c) shows an expanded view up to 130 K.

Figure 3

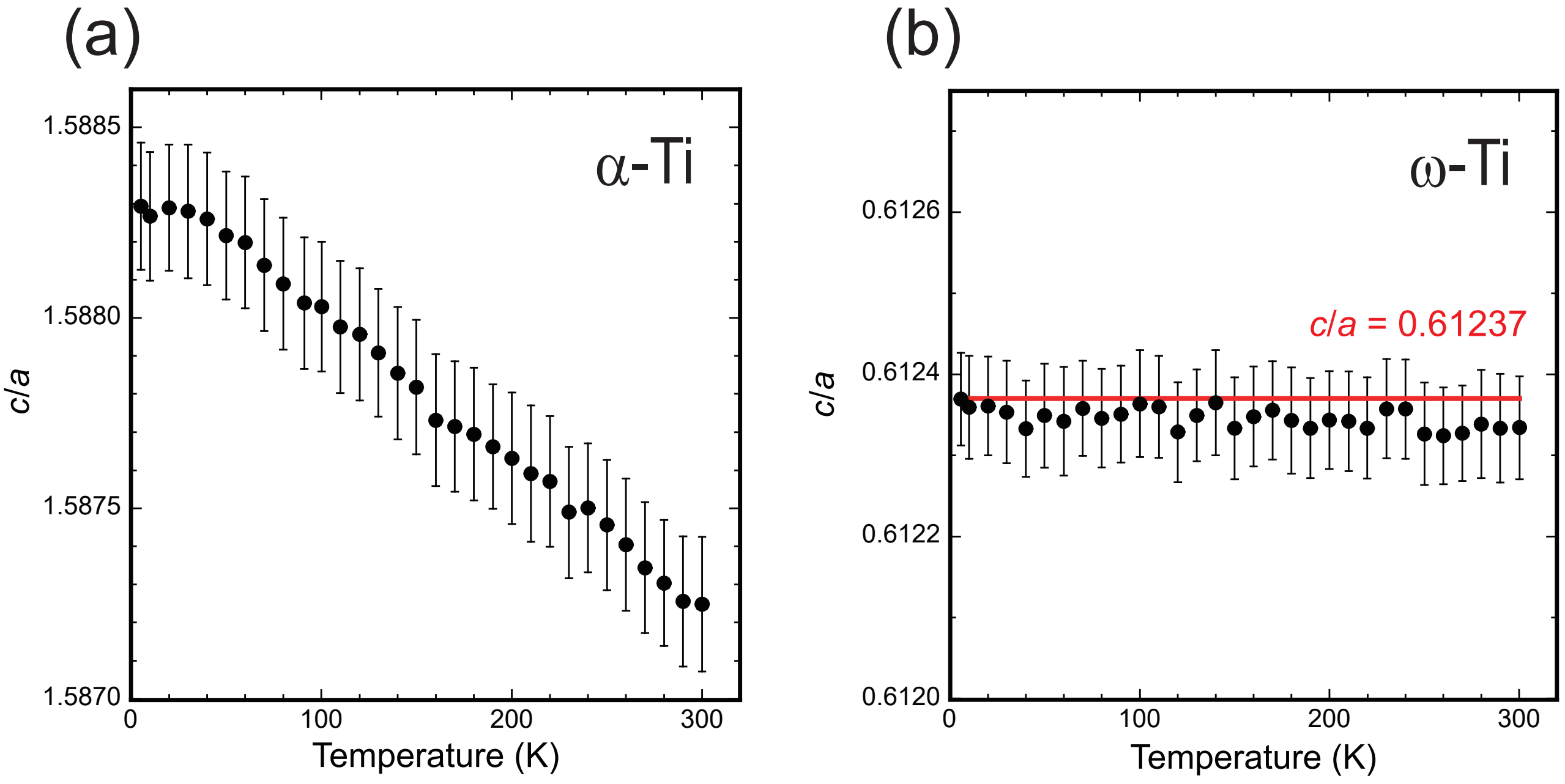

**Fig. 4.** Temperature dependence of the *c*/*a* ratios of (a) α-Ti and (b) ω-Ti. The horizontal line in (b) indicates the ideal bcc-derived value, $\sqrt{3/8} = 0.61237$.

Figure 4

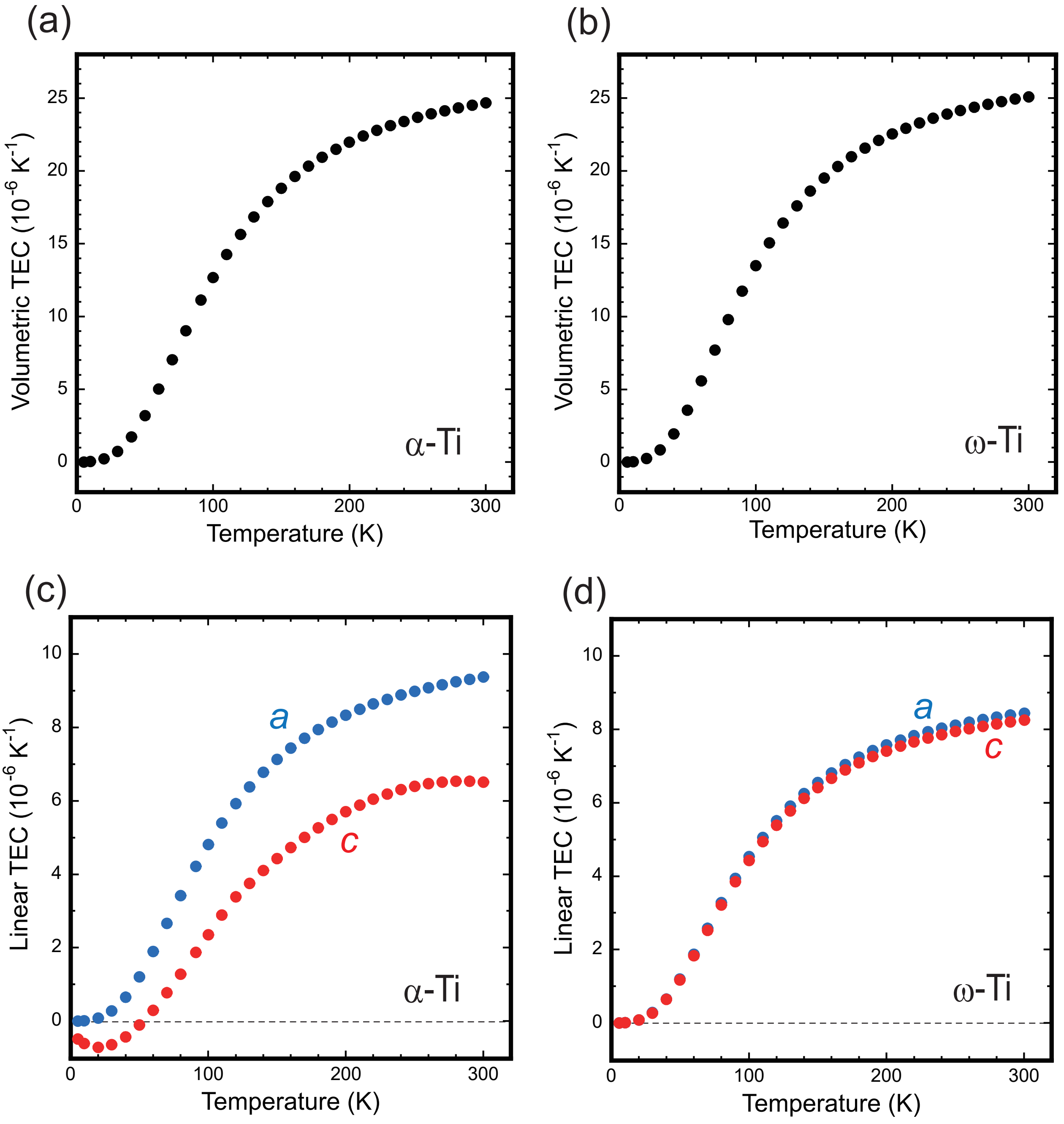

**Fig. 5.** Temperature dependence of thermal expansion coefficients (TECs): (a) volumetric TEC of α-Ti; (b) volumetric TEC of ω-Ti; (c) linear TECs of α-Ti; (d) linear TECs of ω-Ti.

Figure 5

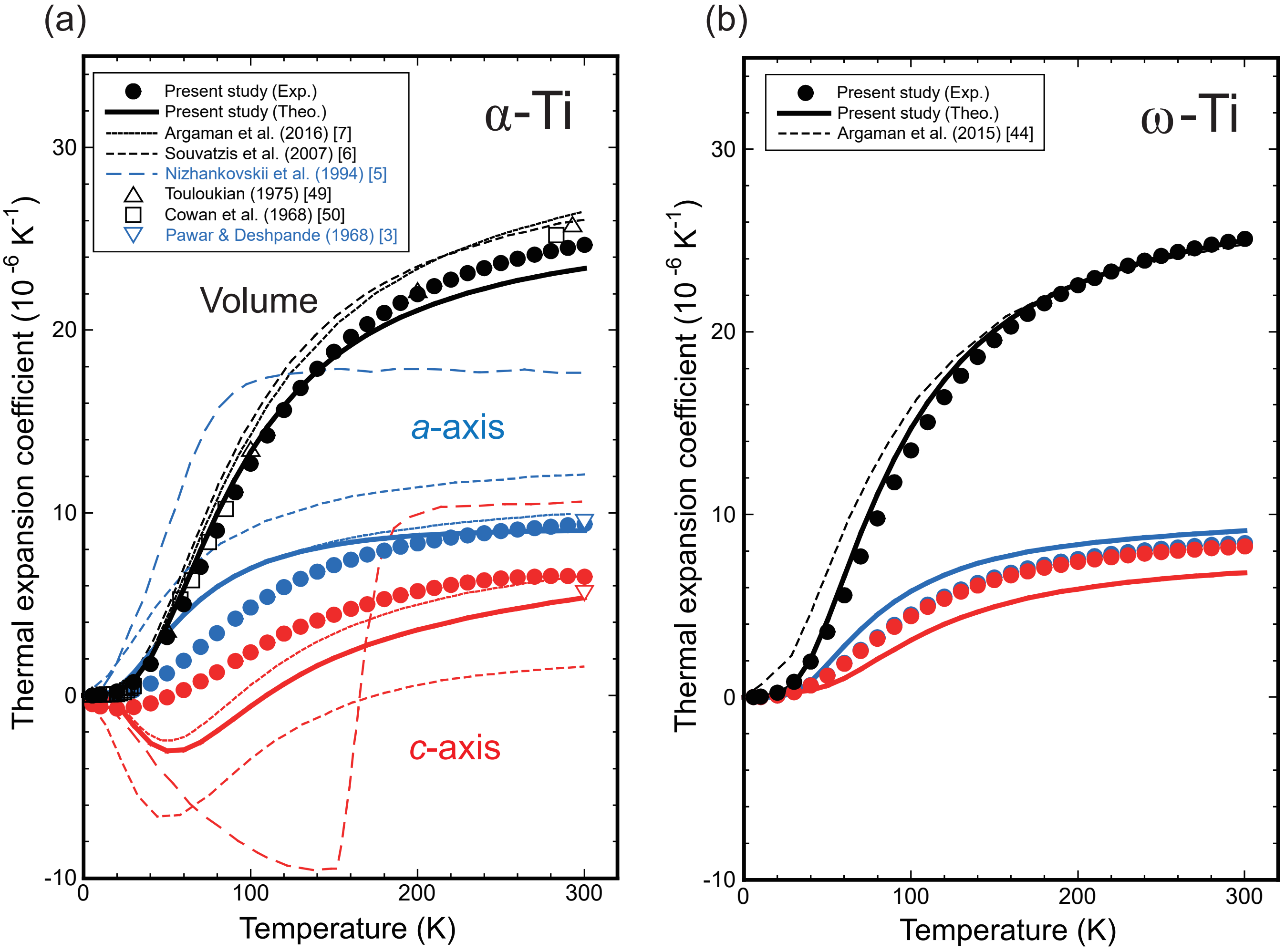

**Fig. 6.** Temperature dependence of the thermal expansion coefficients of (a) α-Ti and (b) ω-Ti determined experimentally and theoretically in the present study. Previously reported experimental and theoretical values are also shown for comparison (see text).

Figure 6

**Table 4**

Axial thermal expansion coefficients and their ratios for α-Ti and ω-Ti at approximately 300 K

| Phase | $\alpha_a$ ($10^{-6}$ $K^{-1}$) | $\alpha_c$ ($10^{-6}$ $K^{-1}$) | $\alpha_c$ /$\alpha_a$ | Remarks |
|---|---|---|---|---|
| α-Ti | 9.37 | 6.51 | 0.69 | *Exp.*, Present study |
| | 9.02 | 5.35 | 0.59 | *Theo.*, Present study |
| | 9.50 | 5.60 | 0.59 | *Exp.*, Pawar & Deshpande [3] |
| | 9.94 | 6.50 | 0.65 | *Theo.*, Argaman et al. [7] |
| | 12.1 | 1.59 | 0.13 | *Theo.*, Souvatzis et al. [6] |
| ω-Ti | 8.44 | 8.25 | 0.98 | *Exp.*, Present study |
| | 9.10 | 6.82 | 0.75 | *Theo.*, Present study |

**Table 5**

Single-crystal elastic constants and elastic anisotropy parameters of α-Ti and ω-Ti obtained in the present study and reported in previous studies

| Phase | $C_{11}$ (GPa) | $C_{33}$ (GPa) | $C_{12}$ (GPa) | $C_{13}$ (GPa) | $C_{44}$ (GPa) | $C_{66}$ (GPa) | $C_{33}/C_{11}$ | $A_{pr}$ | $C_{44}/C_{66}$ | Remarks |
|---|---|---|---|---|---|---|---|---|---|---|
| α-Ti | 170.9 | 190.5 | 89.0 | 76.6 | 42.2 | 40.9 | 1.115 | 0.811 | 1.032 | *Theo.*, Present study |
| | 162.4 | 180.7 | 92.0 | 69.0 | 46.7 | 35.2 | 1.113 | 0.911 | 1.327 | *Exp.*, Fisher & Renken [41] |
| | 172.0 | 190.0 | 82.0 | 75.0 | 45.0 | 45.0 | 1.105 | 0.849 | 1.000 | *Theo.*, Trinkle et al. [52] |
| | 180.2 | 189.1 | 79.0 | 76.3 | 44.6 | 50.6 | 1.049 | 0.823 | 0.881 | *Theo.*, Allen & Bartók [53] |
| ω-Ti | 198.3 | 247.4 | 84.2 | 52.2 | 54.1 | 57.0 | 1.248 | 0.634 | 0.949 | *Theo.*, Present study |
| | 178.9 | 228.4 | 90.2 | 60.8 | 70.5 | 44.4 | 1.277 | 0.987 | 1.588 | *Exp.*, Tane et al. [15] |
| | 194.0 | 245.0 | 81.0 | 54.0 | 54.0 | 56.5 | 1.263 | 0.653 | 0.956 | *Theo.*, Trinkle et al. [52] |
| | 195.0 | 247.5 | 81.1 | 52.7 | 54.4 | 57.0 | 1.269 | 0.646 | 0.954 | *Theo.*, Allen & Bartók [53] |

$A_{pr} = 4C_{44} / (C_{11} + C_{33} - 2C_{13})$;

*Theo.*, values determined by first-principles calculations at 0 K;

*Exp.*, values determined by experimental measurements at approximately 300 K;

The single-crystal elastic constants reported by Tane et al. [15] were estimated from measurements on a polycrystalline specimen prepared by high-pressure torsion.

**Table 6**

Decomposition of the calculated c-axis thermal expansion coefficients of α-Ti and ω-Ti into contributions associated with the a- and c-axis thermal driving forces at selected temperatures. Data are taken from Ref. [38].

| Phase | $T$ (K) | $a$-axis contribution ($10^{-6}$ K$^{-1}$) | $c$-axis contribution ($10^{-6}$ K$^{-1}$) | $\alpha_c$ ($10^{-6}$ K$^{-1}$) |
|---|---|---|---|---|
| α-Ti | 50 | -2.81 | -0.24 | -3.04 |
| | 100 | -7.66 | +7.06 | -0.60 |
| | 300 | -12.84 | +18.19 | +5.35 |
| ω-Ti | 50 | -0.96 | +1.59 | +0.64 |
| | 100 | -3.21 | +6.33 | +3.12 |
| | 300 | -5.56 | +12.38 | +6.82 |

$a$-axis contribution = $2s_{13}\lambda_a$ ;

$c$-axis contribution = $s_{33}\lambda_c$ ;

$\alpha_c = 2s_{13}\lambda_a + s_{33}\lambda_c$ ;

Small differences between the sum of the contributions and the reported $\alpha_c$ values arise from rounding.

## Supplementary Materials

# Low-temperature thermal expansion of the α and ω phases of titanium

Norimasa Nishiyama[1,*], Atsushi Togo[2], Takuya Sasaki[3], Yoshitaka Matsushita[4]

1: Research Center for Materials Nanoarchitectonics, National Institute for Materials Science, 1-1 Namiki, Tsukuba 305-0044, Japan
2: Center for Basic Research on Materials, National Institute for Materials Science, 1-2-1 Sengen, Tsukuba 305-0047, Japan
3: Department of Materials Physics, Graduate School of Engineering, Nagoya University, Furo-cho, Chikusa-ku, Nagoya, Aichi 464-8603, Japan
4: Advanced Engineering and Services Division, National Institute for Materials Science, 1-1 Namiki, Tsukuba 305-0044, Japan

*Corresponding author: Norimasa Nishiyama

Address:
Research Center for Materials Nanoarchitectonics, National Institute for Materials Science, 1-1 Namiki, Tsukuba 305-0044, Japan

Email:
nishiyama.norimasa@nims.go.jp

This Supplementary Material presents the results of Rietveld refinements and the numerical values of the thermal expansion coefficients (TECs) shown in Fig. 6 of the main text. Figure S-1 shows the results of Rietveld refinements of the ω-Ti specimen at 300 and 5.7 K using a single-phase ω-Ti model, together with the result at 300 K using a two-phase ω-Ti + α-Ti model. The refinements were performed using RIETAN-FP [F. Izumi and K. Momma, *Solid State Phenom.*, **130**, 15–20 (2007)]. Tables S-1 to S-4 provide the numerical values of the experimentally and theoretically determined TECs. The TEC values are reported to four decimal places to represent the best-estimate values obtained from the experimental and theoretical analyses; the number of displayed digits does not imply corresponding accuracy or precision.

**Fig. S-1(a)**

A bulk polycrystalline ω-Ti specimen, at 300 K, single-phase model (only ω-Ti).

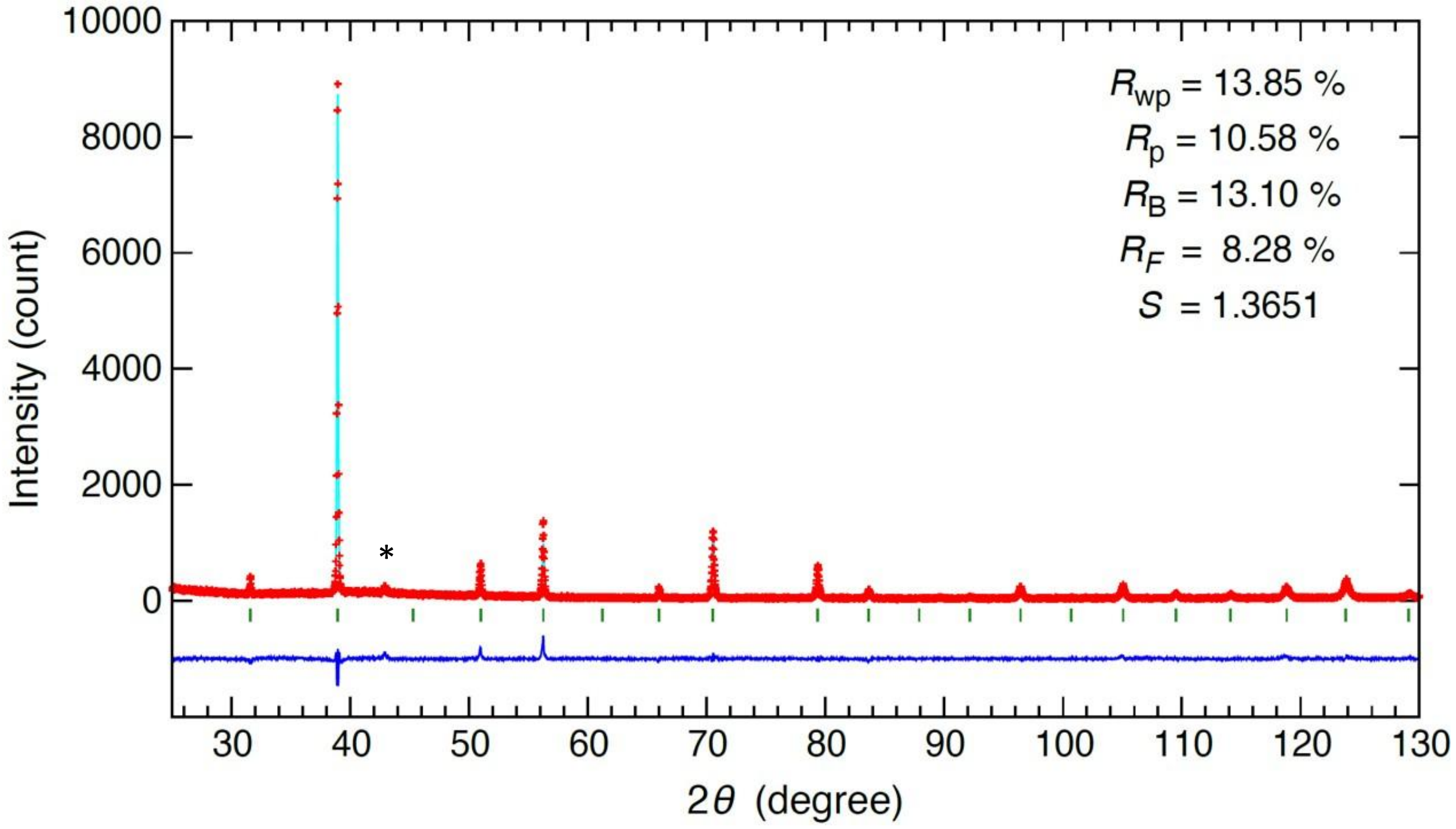

$a$ = 4.6200(4) Å, $c$ = 2.8286(2) Å

**Fig. S-1(b)**

A bulk polycrystalline ω-Ti specimen, at 5.7 K, single-phase model (only ω-Ti).

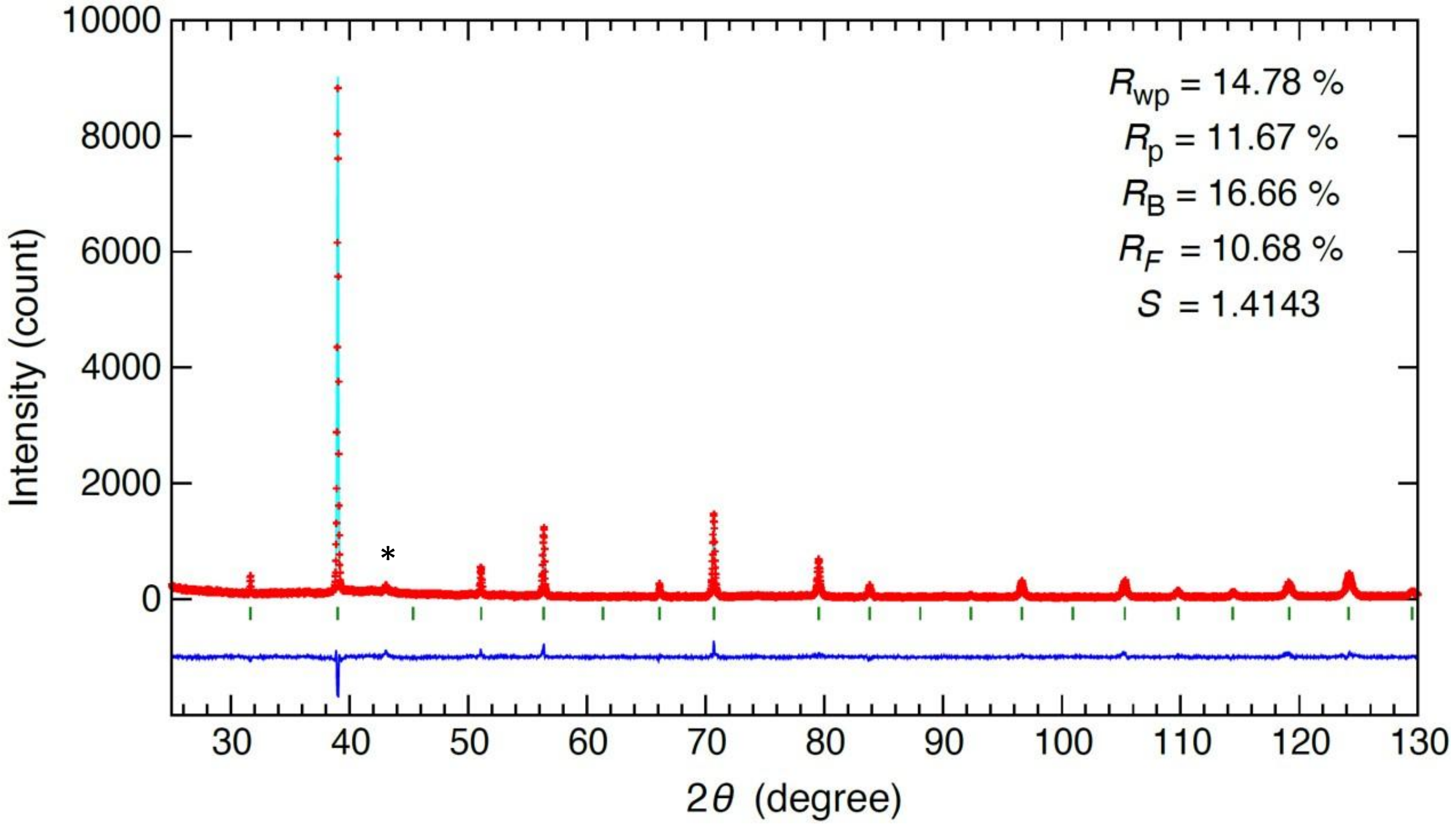

$a$ = 4.6128(3) Å, $c$ = 2.8244(2) Å

**Fig. S-1(c)**

A bulk polycrystalline ω-Ti specimen, at 300 K, two-phase model (α-Ti + ω-Ti).

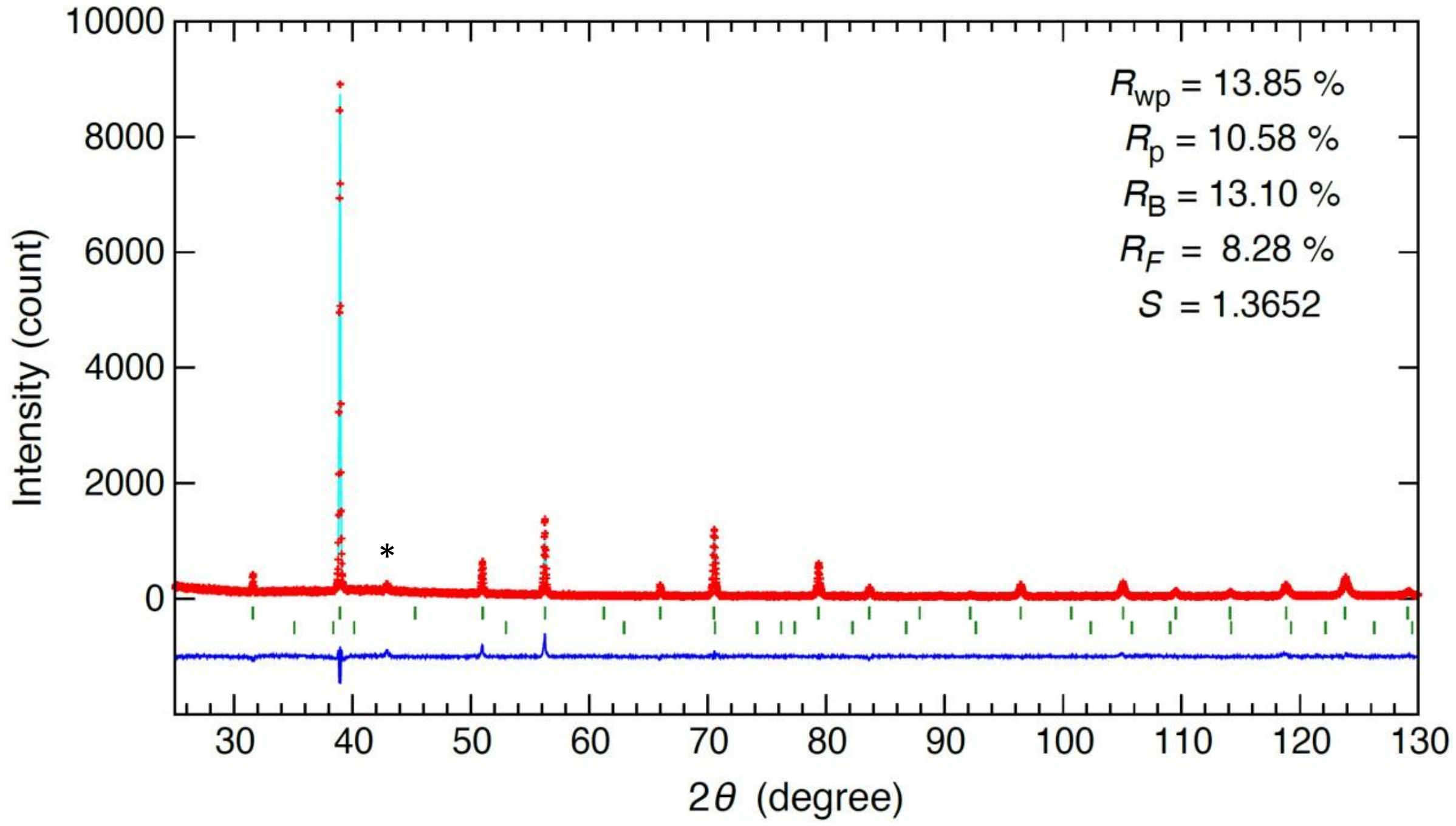

ω-Ti: $a$ = 4.6200(4) Å, $c$ = 2.8286(2) Å

**Fig. S-1(d)**

An enlarged view of Fig. S-1(c) between 25° - 50° of 2$\theta$.

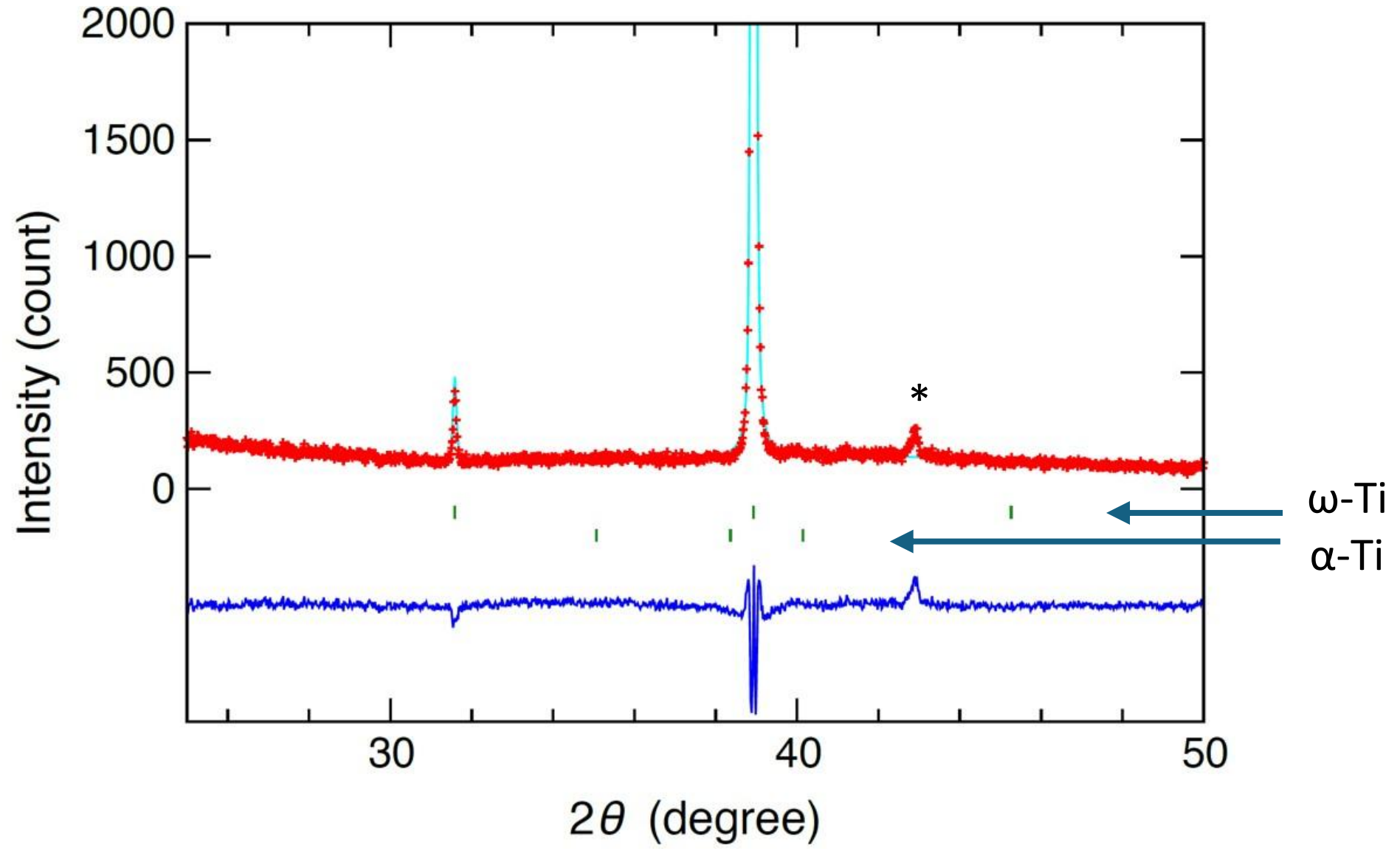

**Fig. S-1.** Summary of Rietveld refinements to assess the phase purity of the ω-Ti specimen. A two-phase ω-Ti + α-Ti model yielded phase fractions of 100.0(1) wt% for ω-Ti and 0.0(1) wt% for α-Ti, without improving the fit. The asterisks indicate the reflections from the Cu sample holder.

**Table S-1**

Experimentally determined thermal expansion coefficients of α-Ti.

| Temp (K) | *a*-axis TEC ($10^{-6}$ $K^{-1}$) | *c*-axis TEC ($10^{-6}$ $K^{-1}$) | *V*-TEC ($10^{-6}$ $K^{-1}$) |
|---|---|---|---|
| 5.2 | 0.0015 | -0.4849 | 0.0038 |
| 9.9 | 0.0101 | -0.6038 | 0.0266 |
| 20.0 | 0.0830 | -0.7099 | 0.2190 |
| 29.9 | 0.2769 | -0.6430 | 0.7309 |
| 40.0 | 0.6523 | -0.4316 | 1.7219 |
| 50.0 | 1.2095 | -0.1086 | 3.1929 |
| 59.9 | 1.8962 | 0.2966 | 5.0056 |
| 70.0 | 2.6627 | 0.7733 | 7.0287 |
| 79.9 | 3.4164 | 1.2810 | 9.0178 |
| 91.0 | 4.2147 | 1.8745 | 11.1241 |
| 100.0 | 4.8063 | 2.3589 | 12.6848 |
| 110.0 | 5.3980 | 2.8861 | 14.2451 |
| 120.0 | 5.9210 | 3.3897 | 15.6239 |
| 130.0 | 6.3800 | 3.7577 | 16.8333 |
| 140.0 | 6.7814 | 4.1053 | 17.8905 |
| 150.0 | 7.1321 | 4.4296 | 18.8135 |
| 160.0 | 7.4387 | 4.7308 | 19.6200 |
| 170.0 | 7.7073 | 5.0087 | 20.3258 |
| 180.0 | 7.9431 | 5.2633 | 20.9449 |
| 190.0 | 8.1506 | 5.4948 | 21.4894 |
| 200.0 | 8.3340 | 5.7030 | 21.9698 |
| 210.0 | 8.4964 | 5.8880 | 22.3949 |
| 220.0 | 8.6408 | 6.0498 | 22.7723 |
| 230.0 | 8.7695 | 6.1884 | 23.1083 |
| 240.0 | 8.8847 | 6.3038 | 23.4084 |
| 250.0 | 8.9880 | 6.3959 | 23.6772 |
| 260.0 | 9.0810 | 6.4649 | 23.9186 |
| 270.0 | 9.1649 | 6.5107 | 24.1361 |
| 280.0 | 9.2409 | 6.5333 | 24.3324 |
| 290.0 | 9.3098 | 6.5328 | 24.5101 |
| 300.0 | 9.3725 | 6.5090 | 24.6714 |

*a*-axis TEC, linear TEC along the *a*-axis;

*c*-axis TEC, linear TEC along the *c*-axis;

*V*-TEC, volumetric TEC.

**Table S-2**

Experimentally determined thermal expansion coefficients of ω-Ti.

| Temp (K) | *a*-axis TEC ($10^{-6}$ $K^{-1}$) | *c*-axis TEC ($10^{-6}$ $K^{-1}$) | *V*-TEC ($10^{-6}$ $K^{-1}$) |
|---|---|---|---|
| 5.7 | 0.0020 | 0.0019 | 0.0058 |
| 10.0 | 0.0105 | 0.0103 | 0.0314 |
| 20.0 | 0.0843 | 0.0824 | 0.2513 |
| 29.9 | 0.2810 | 0.2748 | 0.8379 |
| 39.9 | 0.6537 | 0.6394 | 1.9496 |
| 49.9 | 1.2014 | 1.1751 | 3.5828 |
| 60.0 | 1.8751 | 1.8341 | 5.5920 |
| 69.9 | 2.5811 | 2.5246 | 7.6970 |
| 79.9 | 3.2821 | 3.2103 | 9.7869 |
| 90.0 | 3.9421 | 3.8559 | 11.7540 |
| 100.0 | 4.5316 | 4.4325 | 13.5106 |
| 110.0 | 5.0534 | 4.9429 | 15.0650 |
| 120.0 | 5.5103 | 5.3898 | 16.4254 |
| 130.0 | 5.9081 | 5.7789 | 17.6092 |
| 140.0 | 6.2537 | 6.1170 | 18.6369 |
| 150.0 | 6.5540 | 6.4107 | 19.5293 |
| 160.0 | 6.8152 | 6.6663 | 20.3052 |
| 170.0 | 7.0432 | 6.8892 | 20.9814 |
| 180.0 | 7.2426 | 7.0843 | 21.5724 |
| 190.0 | 7.4177 | 7.2555 | 22.0907 |
| 200.0 | 7.5719 | 7.4064 | 22.5466 |
| 210.0 | 7.7082 | 7.5398 | 22.9491 |
| 220.0 | 7.8292 | 7.6581 | 23.3056 |
| 230.0 | 7.9368 | 7.7634 | 23.6224 |
| 240.0 | 8.0330 | 7.8575 | 23.9048 |
| 250.0 | 8.1192 | 7.9418 | 24.1574 |
| 260.0 | 8.1966 | 8.0175 | 24.3839 |
| 270.0 | 8.2664 | 8.0859 | 24.5876 |
| 280.0 | 8.3296 | 8.1476 | 24.7713 |
| 290.0 | 8.3868 | 8.2036 | 24.9374 |
| 300.0 | 8.4388 | 8.2545 | 25.0879 |

*a*-axis TEC, linear TEC along the *a*-axis;

*c*-axis TEC, linear TEC along the *c*-axis;

*V*-TEC, volumetric TEC.

**Table S-3**

Theoretically determined thermal expansion coefficients of α-Ti.

| Temp (K) | *a*-axis TEC ($10^{-6}$ $K^{-1}$) | *c*-axis TEC ($10^{-6}$ $K^{-1}$) | *V*-TEC ($10^{-6}$ $K^{-1}$) |
|---|---|---|---|
| 0 | 0.0000 | 0.0000 | 0.0000 |
| 10 | 0.0023 | -0.0039 | 0.0007 |
| 20 | 0.3069 | -0.4621 | 0.1517 |
| 30 | 1.2152 | -1.6292 | 0.8012 |
| 40 | 2.3306 | -2.6080 | 2.0531 |
| 50 | 3.4352 | -3.0433 | 3.8270 |
| 60 | 4.4254 | -2.9713 | 5.8794 |
| 70 | 5.2632 | -2.5534 | 7.9729 |
| 80 | 5.9513 | -1.9495 | 9.9530 |
| 90 | 6.5094 | -1.2751 | 11.7438 |
| 100 | 6.9611 | -0.6003 | 13.3219 |
| 110 | 7.3276 | 0.0375 | 14.6927 |
| 120 | 7.6267 | 0.6218 | 15.8752 |
| 130 | 7.8722 | 1.1485 | 16.8928 |
| 140 | 8.0749 | 1.6193 | 17.7692 |
| 150 | 8.2433 | 2.0394 | 18.5260 |
| 160 | 8.3837 | 2.4148 | 19.1821 |
| 170 | 8.5010 | 2.7515 | 19.7536 |
| 180 | 8.5993 | 3.0554 | 20.2541 |
| 190 | 8.6817 | 3.3314 | 20.6948 |
| 200 | 8.7506 | 3.5837 | 21.0849 |
| 210 | 8.8081 | 3.8159 | 21.4321 |
| 220 | 8.8560 | 4.0309 | 21.7429 |
| 230 | 8.8955 | 4.2313 | 22.0223 |
| 240 | 8.9280 | 4.4188 | 22.2748 |
| 250 | 8.9544 | 4.5951 | 22.5039 |
| 260 | 8.9756 | 4.7616 | 22.7127 |
| 270 | 8.9922 | 4.9191 | 22.9036 |
| 280 | 9.0051 | 5.0686 | 23.0788 |
| 290 | 9.0147 | 5.2107 | 23.2401 |
| 300 | 9.0215 | 5.3462 | 23.3891 |

*a*-axis TEC, linear TEC along the *a*-axis;

*c*-axis TEC, linear TEC along the *c*-axis;

*V*-TEC, volumetric TEC.

**Table S-4**

Theoretically determined thermal expansion coefficients of ω-Ti.

| Temp (K) | *a*-axis TEC ($10^{-6}$ $K^{-1}$) | *c*-axis TEC ($10^{-6}$ $K^{-1}$) | *V*-TEC ($10^{-6}$ $K^{-1}$) |
|---|---|---|---|
| 0 | 0.0000 | 0.0000 | 0.0000 |
| 10 | 0.0000 | 0.0101 | 0.0101 |
| 20 | 0.0102 | 0.1352 | 0.1556 |
| 30 | 0.2191 | 0.2563 | 0.6946 |
| 40 | 0.8624 | 0.3909 | 2.1157 |
| 50 | 1.7990 | 0.6423 | 4.2404 |
| 60 | 2.7967 | 1.0385 | 6.6319 |
| 70 | 3.7233 | 1.5364 | 8.9831 |
| 80 | 4.5323 | 2.0768 | 11.1413 |
| 90 | 5.2191 | 2.6131 | 13.0514 |
| 100 | 5.7959 | 3.1177 | 14.7095 |
| 110 | 6.2788 | 3.5772 | 16.1348 |
| 120 | 6.6837 | 3.9877 | 17.3551 |
| 130 | 7.0244 | 4.3506 | 18.3994 |
| 140 | 7.3125 | 4.6697 | 19.2947 |
| 150 | 7.5576 | 4.9496 | 20.0648 |
| 160 | 7.7672 | 5.1955 | 20.7300 |
| 170 | 7.9478 | 5.4117 | 21.3074 |
| 180 | 8.1044 | 5.6026 | 21.8114 |
| 190 | 8.2411 | 5.7717 | 22.2540 |
| 200 | 8.3615 | 5.9221 | 22.6451 |
| 210 | 8.4682 | 6.0565 | 22.9930 |
| 220 | 8.5638 | 6.1771 | 23.3046 |
| 230 | 8.6500 | 6.2858 | 23.5858 |
| 240 | 8.7285 | 6.3843 | 23.8414 |
| 250 | 8.8008 | 6.4737 | 24.0753 |
| 260 | 8.8678 | 6.5554 | 24.2910 |
| 270 | 8.9305 | 6.6302 | 24.4912 |
| 280 | 8.9897 | 6.6989 | 24.6784 |
| 290 | 9.0461 | 6.7623 | 24.8544 |
| 300 | 9.1000 | 6.8209 | 25.0208 |

*a*-axis TEC, linear TEC along the *a*-axis;

*c*-axis TEC, linear TEC along the *c*-axis;

*V*-TEC, volumetric TEC.